\documentclass[prb,reprint,superscriptaddress,amsmath,amssymb,floatfix,twocolumn]{revtex4-1}

\usepackage{graphicx}
\usepackage{epstopdf}
\usepackage{epsfig}
\usepackage{wrapfig}
\usepackage{enumerate}

\usepackage[tableposition=top]{caption}

\DeclareGraphicsExtensions{.ps}
\begin{document}
\title{Analytic continuation-free Green's function approach to\\
correlated electronic structure calculations}
\author{A. \"Ostlin}
\affiliation{Theoretical Physics III, Center for Electronic
Correlations and Magnetism, Institute of Physics, University of
Augsburg, D-86135 Augsburg, Germany}
\author{L. Vitos}
\affiliation{Department of Materials Science and Engineering, Applied Materials Physics,
KTH Royal Institute of Technology, SE-10044 Stockholm, Sweden}
\affiliation{Department of Physics and Astronomy, Division of Materials Theory,
Uppsala University, Box 516, SE-75120 Uppsala, Sweden}
\affiliation{Research Institute for Solid State Physics and Optics, Wigner Research Center for Physics, P.O. Box 49, H-1525 Budapest, Hungary}
\author{L. Chioncel}
\affiliation{Augsburg Center for Innovative Technologies, University of Augsburg,
D-86135 Augsburg, Germany}
\affiliation{Theoretical Physics III, Center for Electronic
Correlations and Magnetism, Institute of Physics, University of
Augsburg, D-86135 Augsburg, Germany}
 
\begin{abstract} 
We present a new charge self-consistent scheme combining Density Functional and Dynamical Mean Field Theory,
which uses Green's function of multiple scattering-type. 
In this implementation the many-body effects are incorporated into the Kohn-Sham iterative scheme without the need
for the numerically ill-posed analytic continuation of the Green's function and of the self-energy. This is achieved
by producing the Kohn-Sham Hamiltonian in the sub-space of correlated partial waves and allows to formulate the Green's
function directly on the Matsubara axis. The spectral moments of the Matsubara Green's function enable us to put together
the real space charge density, therefore the charge self-consistency can be achieved. 
Our results for the spectral functions (density of states) and equation of state curves for transition metal elements,
Fe, Ni and FeAl compound agree very well with those
of Hamiltonian based LDA+DMFT implementations. 
The current implementation improves on numerical accuracy, requires a minimal effort
besides the multiple scattering formulation and can be generalized in several ways that are interesting for applications to real materials.
\end{abstract} 

\maketitle 
\section{Introduction}\label{sec_intro}
Density functional theory (DFT)~\cite{ho.ko.64} in conjunction with the Kohn-Sham scheme~\cite{ko.sh.65} and
the local density approximation (LDA)~\cite{pe.wa.92}, or the generalized gradient approximation (GGA)~\cite{pe.bu.96}, to
the exchange-correlation potential has shown great
success in the computation of ground-state properties of real materials. 
However, the method cannot correctly describe materials where electronic correlations are important,
such as the Mott insulators, $3d$ transition metals and lanthanides. 
One successful approach to improve on the description of the
electronic structure of strongly correlated materials is to merge DFT with
Dynamical Mean Field Theory (DMFT)~\cite{me.vo.89,ge.ko.96,ko.sa.06}. Within DMFT the complicated
many-body lattice problem is mapped self-consistently into a single quantum impurity hybridized with an effective bath. 
Nowadays impurity problems are efficiently solved by various many-body techniques. Hence DMFT
developed into a comprehensive, non-perturbative and thermodynamically consistent theoretical framework
for the investigation of correlated electrons on the lattice. 
The combination of DMFT and DFT, referred to as LDA+DMFT and GGA+DMFT, respectively, has now become the
state-of-the-art method to study correlated materials~\cite{ko.sa.06,held.07}.

During the last decade, various LDA+DMFT implementations have been proposed. 
The early implementations employed a two-step procedure: in the first step the LDA problem was
solved using an effective one-particle Kohn-Sham Hamiltonian and the single-particle wave-functions
(Kohn-Sham basis set) were integrated into the density functional variational approach.
The corresponding Green's function was then obtained using the spectral representation of the
Kohn-Sham Hamiltonian. 
In the second step the interaction problem was treated, i.e., the low-energy effective Hamiltonian was formulated
within a Wannier-like basis obtained through downfolding or, alternatively, by a suitable combination of Kohn-Sham basis sets.
This low-energy Hamiltonian was solved using DMFT. 
Some of the initial LDA+DMFT implementations kept the effective Kohn-Sham potential fixed,
and considered only the self-consistency of the local self-energy. Therefore in these approaches the effect of the self-energy
on the electronic charge was neglected. 
Inserting the self-energy back into the Kohn-Sham iterative scheme allows one to converge towards self-consistency
in both the self-energy and charge. 
Several fully self-consistent Hamiltonian based implementations have been used within the framework of different
basis sets, for example pseudopotential plane waves~\cite{le.bi.08,le.ko.10,amad.12}, linearized muffin-tin
orbitals~\cite{sa.ko.04,po.am.07,gr.ma.07,ma.mi.09,gr.ma.12} and augmented plane waves~\cite{ha.ye.10}.
These procedures follow partly the spirit of the spectral density functional theory (SDFT) proposed by
Savrasov and Kotliar~\cite{sa.ko.04}, in which a self-consistent solution of the Dyson equation is sought.
This leads to a quasiparticle Schr\"odinger (or Dirac) equation with a non-hermitian part in the Hamiltonian.

An elegant way to avoid the difficulties involved in dealing with the non-hermitian Hamiltonian in the SDFT formulation
of LDA+DMFT, is provided by the multiple scattering method based on Green's functions.
Green's function methods have the attractive feature that they can be easily used to treat systems such as surfaces,
defects and impurities~\cite{gu.je.83,sk.ro.91}. 
They can also be employed in connection with the coherent potential approximation (CPA) to study
substitutional disorder~\cite{soven.67}. Common to many Green's function methods is the problem that the electronic
eigenvalue problem is formulated as an \emph{energy-dependent} secular equation, from which it is difficult
to extract the energy bands. 
Therefore, the charge density and the total energy, the relevant quantities for the DFT calculation,
are obtained by integration of the Green's function along contours in the complex energy plane~\cite{ze.de.82}. 
Some of the first charge self-consistent implementations of LDA+DMFT with a Green's function formulation
of the Kohn-Sham DFT were implemented within the exact muffin-tin orbitals (EMTO)
method~\cite{ch.vi.03} and the Korringa-Kohn-Rostoker (KKR) method~\cite{mi.ch.05}. 

One of the major goals of any self-consistent LDA+DMFT computation is to answer the question of how
the effects of electronic correlation modify the equilibrium properties, like lattice parameters and
bulk modulus, beyond the LDA. It is hence necessary to calculate accurate total energies within LDA+DMFT,
from which the equilibrium quantities can be derived. Several of the ground state quantities and spectral
properties have already been discussed~\cite{le.po.11,le.an.14} within the Hamiltonian framework.
Despite the many successes of Green's function-based LDA+DMFT methods~\cite{minar.11}, several numerical
difficulties still remain for total energy calculations. 
When the Green's function based LDA+DMFT scheme is executed in practice, Pad\'e approximants~\cite{vi.se.77}
(rational polynomials) are used to pass Green's functions from the complex energy contour to the Matsubara frequencies,
and to return with the self-energy from the Matsubara frequencies back to the complex contour. Besides being sensitive
to numerical noise~\cite{be.go.00}, Pad\'e approximants may miss important features, that can only be captured by resummation
of the continued fraction to infinite order~\cite{wall.48,bake.75,be.go.00}.
In recent years some methods have been proposed in order to improve on the original
Pad\'e approximation technique~\cite{os.ch.12,sc.lo.16,no.sc.16} to the analytic continuation of the Green's function,
but as of yet no fully satisfactory solution to this problem exists.
Such numerical problems are presently a bottleneck for an accurate and stable self-consistent Green's
function based LDA+DMFT method that can produce reliable total energies. 

The success of LDA+DMFT consists in its ability to produce a self-consistent, numerically manageable
approximation for the spectral function and for lattice properties at equilibrium. It is desirable that
LDA+DMFT developments be exact {\it in principle}, and that even approximate perturbative solvers should
give good results, irrespective of whether a Hamiltonian or Green's function method is used.
For these reasons it is essential to pursue alternative methods that improve on the numerical accuracy. In general,
for a Green's function formulation of the LDA+DMFT the knowledge of the non-interacting Green's
function along the imaginary axis is required. 
Consequently, our primary objective of the present paper is to describe an approach which yields an accurate
Green's function in Matsubara frequencies which can be used in the DMFT part and, at the same time,
in constructing the charge density. 

Our novel method makes the analytic continuation during the self-consistent Kohn-Sham iterations unnecessary.
The key observation that triggered this method development is that the charge density is the only
ingredient needed to close the Kohn-Sham self-consistent loop. 
The charge density \emph{difference} between correlated and non-correlated calculations, evaluated on the
imaginary Matsubara-axis, is taken as the correction on the DFT level charge density. Quantities like eigenvalues,
Green's functions and self-energies are only auxiliary quantities in this respect.
In the method, $z$MTO+DMFT, presented here the Green's function in Matsubara frequencies is evaluated from the LMTO eigenstates,
i.e., in the basis of linearized partial waves.
The choice to take the character $z$ in the denomination $z$MTO+DMFT is to remind of the fact
that the Green's functions in DFT are usually computed along a general complex contour mesh, i.e. $G(z)$, for a given muffin-tin potential. 
We implemented this scheme starting from our previous EMTO+DMFT method~\cite{ch.vi.03}, which has been successfully
used to study correlated systems, such as $3d$ transition metals and compounds~\cite{ch.vi.03,ch.al.07,ka.ir.08},
magnetic heterostructures~\cite{be.ho.11} and transport properties through layered structures~\cite{ch.mo.15}.
The use of a Green's function method opens the possibility to study systems that deviate from
perfect crystalline conditions, such as alloys and surfaces.

The paper is organized as follows: Sec.~\ref{sec_muffin} gives an overview of the muffin-tin formalism
for the solution of the Kohn-Sham equations. Sec.~\ref{sec_impl} presents the new charge-self-consistent
implementation, followed by results in Sec~\ref{sec_res}. A conclusion and outlook is given in Sec.~\ref{sec_conc}.

\section{Overview of the muffin-tin formalism}\label{sec_muffin}

Muffin-tin based methods have in common that they partition space 
into spherical \emph{muffin-tins}, centered around the ions in the lattice, and the \emph{interstitial},
the area outside of the muffin-tins. Inside the muffin-tins the effective potential is assumed to be spherically symmetric,
while it is taken to be constant in the interstitial. 
The Kohn-Sham equations are solved separately within these regions, and the solution for the entire space is
found by imposing boundary conditions between the muffin-tins and the interstitial. The algebraic
formulation of the matching conditions takes the form of a secular equation, which is in general
energy-dependent. Sec.~\ref{subsec_emto} describes this concept for the EMTO method. 
Sec.~\ref{subsec_lemto} briefly reviews the concept of basis function linearization,
which is important for the construction of the correlated orbitals in this work.

\subsection{Charge density and the complex contour Green's function in the EMTO basis set}\label{subsec_emto}

Within the muffin-tin formalism, the effective Kohn-Sham potential $V^\sigma_{eff}({\bf r})$
($\sigma$ denotes the spin) in the single-electron Kohn-Sham equations, labeled by state index $j$,
\begin{equation} \label{kosh}
\left[\nabla^2-V^\sigma_{eff}({\bf r})\right] \Psi^\sigma_{j}({\bf r})
\;=\;\epsilon^\sigma_j\Psi^\sigma_j({\bf r}),
\end{equation}
is approximated by spherical muffin-tin wells
centered at lattice sites $\mathbf{R}$. The exchange-correlation part of $V^\sigma_{eff}({\bf r})$
will in the following always be approximated by the spin-polarized LDA, and we will from now
on suppress the spin index for simplicity. For the EMTO basis set~\cite{an.je.94.2,vi.sk.00,vito.01,vitos.10}, 
the one-electron wave-functions
are expanded in exact muffin-tin orbitals $\bar\psi^a_{RL}$,
\begin{equation}\label{emto}
\Psi_j({\bf r}) = \sum\limits_{RL} \bar\psi^a_{RL}({\epsilon_j,{\bf r}_R})v^a_{RL,j},
\end{equation}
where $L\equiv(l,m)$ denotes the orbital and azimuthal quantum numbers respectively, and ${\bf r}_R\equiv r_R\hat{r}_R =
{\bf r}-{\bf R}$, where the vector notation for the index $R$ has been omitted. The superscript $a$ denotes the
screening parameter. The orbitals $\bar\psi^a_{RL}$ are linear combinations of partial
waves $\phi^a_{LR}(r_R)$, which are normalized solutions of the radial Schr\"odinger eqution
inside the muffin-tins with spherical potential $V_{eff}(r_R)$,
\begin{equation}\label{radialsch}
\frac{\partial^2 r_R \phi_{Rl}(z,r_R)}{\partial r^2_R} = \left[ \frac{l(l+1)}{r^2_R} +
V_{eff}(r_R) - z\right]r_R \phi_{Rl}(z,r_R),
\end{equation}
and of the solutions in the interstitial region~\cite{vitos.10}. The angular momentum sum in Eq. (\ref{emto}) can
in practice be truncated at $l_{max}=3$, making the basis minimal. Since the orbitals
are centered around the lattice sites $R$, the basis is ``local'', making
it suitable as a basis for correlated orbitals within DMFT. The coefficients
$v^a_{RL,j}$ are determined from the condition that the expansion should
fulfill Eq. (\ref{kosh}) in all space, i.e. the orbitals should be everywhere continuous and
have no derivative discontinuities (kinks) anywhere. In the EMTO formalism this leads to the 
kink cancellation equation:
\begin{equation}\label{kink_v}
K^a_{RL,R'L'}(\epsilon_j)v^a_{RL,j} = 0
\end{equation}
which is equivalent to the KKR tail cancellation equation~\cite{vitos.10}, written in a screened representation.
The quantity $K^a_{RL,R'L'}(\epsilon_j)$ defines the kink matrix for an arbitrary complex energy $z$ and has the form:
\begin{equation}\label{g2}
K^{a}_{RL,R'L'}(z) \equiv
a\delta_{RR'}\delta_{LL'}D^{a}_{RL}(z) - aS^{a}_{RL,R'L'}(z).
\end{equation}
$D^{a}_{RL}(z)$ denotes the EMTO logarithmic derivative function
\cite{vi.sk.00,vito.01}, and $S^{a}_{RL,R'L'}(z)$ is the slope matrix \cite{an.sa.00}.
Note that Eq.~(\ref{kink_v}) is an \emph{energy-dependent} secular equation, which allows one to
determine the eigenvalues $\epsilon_j$. These are obtained using numerical search algorithms
for the roots of the secular determinant along the real energy
axis. To simplify the notation further, we suppress the index for the screening parameter $a$.

For translation invariant systems, the index $R$ runs over the atoms in the primitive cell only, 
and the Fourier transformation of Eq.\ (\ref{g2}) produces a matrix equation in the reciprocal space:
\begin{eqnarray}\label{g2_k}
\sum_{R''L''}K_{R'L',R''L''}({\bf k},z)g_{R''L'',RL}({\bf k},z) = \delta_{R'R} \delta_{L'L}
\end{eqnarray}
that is solved using Green's function methods. Accordingly, the path operator $g_{R''L'',RL}({\bf k},z)$ is 
the unique solution of Eq.~(\ref{g2_k}) (the inverse of the kink matrix $K_{R'L',R''L''}({\bf k},z)$) that 
fulfills the combination of lattice symmetry and boundary conditions. The elements of the kink matrix are 
constructed from the Bloch wave vector (${\bf k}$) dependent slope matrix~\cite{vitos.10}. 
Since the energy derivative of the kink matrix, ${\dot K}_{RL,R'L'}({\bf k},z)$,
gives the overlap matrix for the EMTO basis set~\cite{an.sa.00}, these are used to 
normalize the path operator  $g_{R''L'',RL}({\bf k},z)$ and construct 
the matrix elements of the EMTO Green's function~\cite{vi.sk.00,vito.01}
\begin{eqnarray}\label{green}
G_{RL,R'L'}({\bf k},z)\;&=&\;
\sum_{R''L''}g_{RL,R''L''}({\bf k},z){\dot K}_{R''L'',R'L'}({\bf k},z)\nonumber\\
&&-\delta_{RR'}\delta_{LL'} I_{RL}(z),
\end{eqnarray}
where $I_{RL}(z)$ accounts for the unphysical poles of ${\dot K}_{RL,R'L'}(z)$~\cite{vito.01,vitos.10}.
The total number of states at the Fermi level $E_F$ is obtained as
\begin{equation}\label{nos}
N(E_F) = \frac{1}{2\pi i}\sum_{RL}\oint
\sum\limits_{\mathbf{k}}G_{RL,RL}({\bf k},z)\;\mathrm{d}{\bf k}\;\mathrm{d}z,
\end{equation}
where the energy integral is carried out on a complex contour that cuts the real axis 
below the bottom of the valence band and at $E_F$. 
The \textbf{k}-summation is performed over the Brillouin zone (BZ). 

To close the Kohn-Sham self-consistency scheme
requires the computation of the charge density. Within the 
EMTO method this is achieved through the real space
path operator (corrected for unphysical poles similarly 
as in Eq.~(\ref{green})~\cite{vitos.10}) integrated over  
the same complex contour that is used to determine $E_F$,
\begin{eqnarray}\label{emtocharge}
&n&({\bf r})=\sum\limits_R n_R({\bf r}_R); \quad n_R({\bf r}_R) = \sum\limits_L n_{RL}(r_R)Y_L(\hat{r}_R),\nonumber\\
&n&_{RL}(r_R) = \frac{1}{2\pi i} \oint \times\nonumber\\ &&\sum_{L',L''} C_{LL'L''}
Z_{Rl''}(z,r_R) g_{RL'',RL'}(z) Z_{Rl'}(z,r_R) \;\mathrm{d}z,
\end{eqnarray}
where $C_{L'LL''}$ is a real Gaunt number. Eq.~(\ref{emtocharge}) is valid within the
muffin-tin spheres and for $l\leq l_{max}$, and $Z_{Rl}(z,r_R)=N_{Rl}(z)\phi_{Rl}(z,r_R)$,
where $N_{Rl}(z)$ is a normalization function~\cite{vito.01,vitos.10}. 
The specific set of real harmonics is denoted by $Y_L(\hat{r}_R)$.

\subsection{Charge density and the Matsubara Green's function in the LMTO basis set}\label{subsec_lemto}
An alternative solution of Eq.~(\ref{kosh}) is obtained by the linearized muffin-tin orbitals
(LMTO)~\cite{ande.75,an.je.86} method. The same muffin-tin shape is used for the potentials as in
the EMTO method, but with the additional approximation that the interstitial region is neglected,
leading to the atomic sphere approximation (ASA). 
The LMTOs $\chi^{\gamma}_{RL}$ are constructed from the partial wave solutions $\phi_{Rl}$ inside
the muffin-tin spheres, computed at an arbitrary energy $\epsilon_{Rl\nu}$ (commonly chosen as the
center of gravity of the occupied part of the band), and from
the energy derivative of the partial wave, 
$\dot{\phi}_{Rl} =\partial \phi_{Rl}/ \partial \epsilon |_{\epsilon=\epsilon_{Rl\nu}}$, viz.
\begin{equation}\label{lmto}
\chi^{\gamma}_{RL}(\mathbf{r}_R) = \phi_{Rl}(\mathbf{r}_R) + \sum\limits_{R'L'}
\dot{\phi}_{R'l'}(\mathbf{r}_R)h^{\gamma}_{R'L',RL}(\mathbf{k}).
\end{equation}
The omitted energy argument of the partial wave $\phi_{Rl}$ means that the function is evaluated at an energy $\epsilon_{Rl\nu}$. 
In Eq.~(\ref{lmto}), $h^{\gamma}_{R'L',RL}(\mathbf{k})$ is defined as
\begin{equation}
h^{\gamma}_{R'L',LR}(\mathbf{k}) \equiv H^{\gamma}_{R'L',LR}(\mathbf{k}) - 
\epsilon_{Rl\nu}\delta_{L'L}\delta_{R'R},
\end{equation}
where $H^{\gamma}_{R'L',RL}(\mathbf{k})$ is the Kohn-Sham Hamiltonian in the so-called nearly
orthogonal $\gamma-$representation~\cite{an.je.84,an.je.86} viz.
\begin{equation}\label{hamiltonian}
H^{\gamma}_{RL,R'L'}(\mathbf{k}) = C_{Rl}\delta_{L'L}\delta_{R'R}+
\sqrt{\Delta_{Rl}}S^{\gamma}_{RL,R'L'}(\mathbf{k})\sqrt{\Delta_{R'l'}},
\end{equation}
where $S^{\gamma}_{RL,R'L'}$ are the LMTO structure constants, and the potential parameters 
$C_{Rl}$ and $\Delta_{Rl}$ are computed from the partial waves $\phi_{Rl}$ according to the prescription given in Ref.~\onlinecite{an.je.86}.
With the energy independent LMTO basis functions, Eq.~(\ref{lmto}), the lattice wave function (i.e. the linear muffin-tin wave function): 
\begin{equation}\label{latt_lmto}
\Psi_j({\bf r}) = \sum\limits_{RL} \chi^\gamma_{RL}({{\bf r}_R})u_{RL,j},
\end{equation}
follows the \emph{energy-independent} eigenvalue problem: 
\begin{equation}\label{h_lmto}
H^{\gamma}_{R'L',RL}(\mathbf{k})u_{RL,j}(\mathbf{k})=\epsilon_j(\mathbf{k})u_{RL,j}(\mathbf{k}), 
\end{equation}
where the Hamiltonian eigenvalues $\epsilon_j(\mathbf{k})$
provides the band structure, and the eigenvectors $ u_{RL,j}(\mathbf{k}) $ contain Bloch vector specific information. 

\subsubsection{Moments from the LMTO eigenstates and complex contour}

Once the LMTO Hamiltonian has been diagonalized, Eq.~(\ref{h_lmto}), the energy moments can be evaluated as
\begin{equation}\label{lmtomom}
{\mathcal M}^q_{Rl} \equiv
\sum\limits^{occ.}_{j\mathbf{k}} [\epsilon_j(\mathbf{k})-\epsilon_{Rl\nu}]^q \sum\limits_L |u_{RL,j}(\mathbf{k})|^2,
\end{equation}
where the $q=0$ and $q=1$ moments correspond to the orbitals occupation and one-electron energies, respectively.
Note that the moments computed with the help of Eq.~(\ref{lmtomom}), is along the real energy axis.

To make contact with DMFT we point out that the LMTO method has been already used to construct Green's functions:
either from the potential parameters directly or 
from the Lehmann (eigenvalue) representation~\cite{an.je.86,sk.ro.91,po.am.07}:
\begin{eqnarray}\label{lmtogreen}
G_{RL,R'L'}(z) = \sum\limits_{j\mathbf{k}} \frac{u_{RL,j}(\mathbf{k})[u_{R'L',j}(\mathbf{k})]^{\dagger}}{z-\epsilon_j(\mathbf{k})}.
\end{eqnarray}
The energy moments can then be computed along a similar complex contour as in the EMTO method~\cite{sk.ro.91},
using the site and orbital diagonal part of the Green's function,
$(R'L') \equiv (RL)$ viz.
\begin{eqnarray}\label{glmtocontour}
{\mathcal M}^{q}_{Rl} = \frac{1}{2 \pi i} \oint \sum\limits^{l}_{m=-l} (z-\epsilon_{Rl\nu})^{q} G_{RL,RL}(z) \mathrm{d}z,
\end{eqnarray}
where we remind the reader of the definition $L\equiv(l,m)$. 
The eigenvalue summation done in Eq.~(\ref{lmtomom}), is now replaced with the complex contour integration Eq.~(\ref{glmtocontour}).
The knowledge of the moments and the partial waves allows the computation of the charge density~\cite{an.je.86}, viz.
\begin{eqnarray}\label{lmtocharge}
n_{Rl}(r_R) &=& {\mathcal M}^0_{Rl} |\phi_{Rl}(r_R)|^2 + {\mathcal M}^2_{Rl} |\dot{\phi}_{Rl}(r_R)|^2 \nonumber\\
&&+ 2{\mathcal M}^1_{Rl} \phi_{Rl}(r_R)\dot{\phi}_{Rl}(r_R)  \nonumber\\
&&+ {\mathcal M}^2_{Rl} \phi_{Rl}(r_R)\ddot{\phi}_{Rl}(r_R),  
\end{eqnarray}
and the  DFT self-consistency loop can be closed.

Note that one advantage of the LMTO Green's function over a multiple-scattering Green's function
is that its spectrum is discrete and upwards bound, i.e. it does not contain the free-electron continuum~\cite{wein.90}.

\subsubsection{Moments from Matsubara LMTO Green's function}

Eq.~(\ref{lmtogreen}) can be also defined for the Matsubara frequencies
$i\omega _{n}=(2n+1)i\pi T$, where $n=0,\pm1,...$, and $T$ is the temperature.
Pourovskii et al.~\cite{po.am.07}, showed recently that the LMTO zeroth energy moments can be extracted
also from the imaginary frequency domain by standard Matsubara summation~\cite{fetter.walecka}, viz.
\begin{equation}\label{0mom}
{\mathcal M}^0_{Rl} = T\sum\limits_n \sum\limits^{l}_{m=-l} \sum\limits_{\mathbf{k}} G_{RL,RL}(\mathbf{k},i\omega_n) 
e^{i\omega_n 0^+}.
\end{equation}
with the $\mathbf{k}$-resolved Green's function given by the Lehmann representation
\begin{eqnarray}\label{lmtogreenmats}
G_{RL,R'L'}(\mathbf{k}, i\omega) = \sum\limits_{j} \frac{u_{RL,j}(\mathbf{k})[u_{R'L',j}(\mathbf{k})]^{\dagger}}{i\omega_n+\mu-\epsilon_j(\mathbf{k})}.
\end{eqnarray}
The local Green's function is computed as:
\begin{eqnarray}\label{lmto_loc_green}
G_{RL,R'L'}(i\omega) = \sum_{{\bf k}} G_{RL,R'L'}(\mathbf{k}, i\omega).
\end{eqnarray}
The higher order moments can be calculated as products of the zeroth order moment $\mathcal{M}^0_{Rl}$,
and $\epsilon_j(\mathbf{k})-\epsilon_{Rl\nu}$,
\begin{eqnarray}\label{hmom}
{\mathcal M}^1_{Rl} &=& \sum\limits^{occ.}_{j\mathbf{k}} {\mathcal M}^0_{Rl} [\epsilon_j(\mathbf{k})-\epsilon_{Rl\nu}] \nonumber \\
{\mathcal M}^2_{Rl} &=& \sum\limits^{occ.}_{j\mathbf{k}} {\mathcal M}^0_{Rl} [\epsilon_j(\mathbf{k})-\epsilon_{Rl\nu}]^2,
\end{eqnarray}
The charge density can be computed again from Eq.~(\ref{lmtocharge}).
Note that a cutoff at a finite frequency will lead to inaccurate Matsubara sums~\cite{ni.st.94}. This can be
corrected to some extent by taking the analytic tail of the Green's function into account~\cite{po.ka.05,po.am.07}.

\subsection{Incorporating the local many-body self-energy}\label{subsec_dmft}
After the brief review of the energy-dependent and the energy-linearized basis sets we proceed
with discussing a combination of these methods which allows to include the local DMFT self-energy in a charge
self-consistent way. The DMFT maps self-consistently the many-body lattice problem to an impurity model,
which can be solved by various many-body techniques and produces the impurity Green's function and the
local self-energy~\cite{ko.sa.06}. The DMFT self-consistency condition is obtained by imposing that the impurity
Green's function is the same as the lattice local Green's function. 

\begin{figure}
	\includegraphics[scale=1]{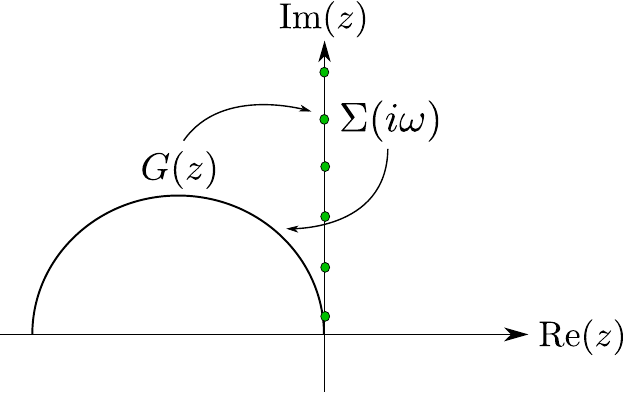}
\caption{(Color online) Schematic picture of the complex energy contour and the Matsubara frequencies
used in the EMTO+DMFT method~\cite{ch.vi.03}. Two Pade continuations are needed, Eqs.~(\ref{pade1}) and (\ref{pade2}),
which are numerically ill-posed problems.}
	\label{complexplane}
\end{figure}

In the EMTO+DMFT method \cite{ch.vi.03}, the self-consistent procedure starts with a guess for the local
self-energy $\Sigma_{RL,RL'}(z)$ to be combined, through the Dyson equation, with the ${\bf k}$-resolved
LDA Green's function, Eq.~(\ref{green}), which represents the ``non-interacting'' lattice Green's function: 
\begin{eqnarray}  \label{G0}
\left[G_{RL,R'L'}({\bf k},z)\right]^{-1} & = &
\left[G_{RL,R'L'}^{{\rm LDA}}({\bf k},z)\right]^{-1} -
\delta_{RR'}\Sigma_{RL,RL'}(z), \nonumber \\
G_{RL,R'L'}(z) &=& \sum\limits_{\mathbf{k}}G_{RL,R'L'}({\bf k},z),
\end{eqnarray}
The local Green's function is extracted from Eq.~(\ref{G0}) on the complex contour:
$G_{RL,RL'}(z)$. Its matrix elements are analytically continued to the Matsubara frequencies:
\begin{equation}
G_{RL,RL'}(z) \xrightarrow{\mathrm{Pad\acute{e}}} G_{RL,RL'}(i\omega). \label{pade1}
\end{equation}
In the next step one has to construct the bath Green's function which specifies the impurity problem, which
within EMTO+DMFT is computed from the analytically continued lattice local Green's function and the self-energy:
\begin{equation} \label{Gtilde}
\left[{\cal G}_{RL,R'L'}(i\omega)\right]^{-1} =
\left[G_{RL,R'L'}(i\omega)\right]^{-1}+
\delta_{RR'}\Sigma_{RL,RL'}(i\omega).
\end{equation} 
The many-body problem is solved on the Matsubara axis, and the resulting self-energy is then analytically
continued to the semi-circular contour:
\begin{equation}
\Sigma_{RL,RL'}(i\omega) \xrightarrow{\mathrm{Pad\acute{e}}} \Sigma_{RL,RL'}(z), \label{pade2}
\end{equation} 
in order to close the LDA+DMFT loop. 
In Figure \ref{complexplane} we illustrate the contours used in the EMTO+DMFT calculations. Accordingly,
the self-consistency procedure requires two Pad\'{e} analytic continuation~\cite{vi.se.77,ch.vi.03} steps,
that has to be controlled numerically.

In order to close the charge self-consistent loop, the LDA+DMFT path operator $g_{RL,R''L''}(z)$ is extracted from
the interacting Green's function~(\ref{G0}), while the real-space charge density is computed according
to Eq.~(\ref{emtocharge}) substituting the LDA path operator with the corresponding LDA+DMFT  path operator.
The new effective Kohn-Sham potential is obtained by solving the Poisson equation, and the scheme is
iterated until self-consistency is achieved.

The LMTO method has previously been used as a choice for charge self-consistent basis sets. 
In particular, Pourovskii et al.~\cite{po.am.07} implemented an LDA+DMFT scheme
in the LMTO-ASA method. In the case of LMTO-ASA, the LDA level Green's function is easily evaluated along
the imaginary axis (Eq.~(\ref{lmtogreenmats})), and the self-energy $\Sigma(i\omega)$ is embedded via
the Dyson equation to obtain the LMTO LDA+DMFT level Green's function. After performing the \textbf{k}-sum,
the bath Green's function is given similarly as in Eq.~(\ref{Gtilde}), and is given as input to the DMFT
impurity solver. In order to close the charge self-consistent loop, the energy moments are computed
as in Eqs.~(\ref{0mom}-\ref{hmom}), with the exception that the Green's function in Eq.~(\ref{0mom}) is 
now on the LDA+DMFT level. The charge density is then computed from the energy moments as outlined
in Eq.~(\ref{lmtocharge}).

\section{Implementation of the new $z$MTO+DMFT method}\label{sec_impl}

\subsection{Motivation}
In this Section, we present a novel scheme that removes the need for the ill-posed analytic
continuations Eqs.~(\ref{pade1}) and (\ref{pade2}), during the self-consistent loops. 
Two main ideas are used to achieve this: (i) the Green's function can be well approximated
by linearization of the muffin-tin orbitals, and (ii) the charge density can be calculated by Matsubara summation. 

\subsubsection{Elimination of $G(z) \rightarrow G(i\omega)$:\\ the benefit of a linearized basis set} \label{g_pade}
A major difference between the Green's function within EMTO, Eq.~(\ref{G0}), and within LMTO-ASA,
Eq.~(\ref{lmtogreen}), is that the latter can be easily evaluated for any energy once the potential parameters are known.
The EMTO Green's function on the other hand requires the computation of the slope matrix and the
solution of the radial Schr\"odinger equation at each energy point along the complex contour, and this is a numerically demanding task. 
The two Green's functions should be equivalent up to the error in the linearization imposed on
the kink cancellation condition~\cite{ta.ar.00}, reflecting the error of the linearization of the muffin-tin basis set.

Based on the formal equivalence of these methods, and the similar results for the corresponding quantities
(Green's functions and moments of these), we propose the following: 
\begin{itemize}
\item The EMTO Green's function should be used for LDA  calculations, 
\item The LMTO Green's function should be used for DMFT calculations.
\end{itemize}
This replaces the need of a Pad\'e approximant with a linearization of the basis set, a more well controlled approximation.

To be specific, we outline the procedure: At each Kohn-Sham iteration, the kink matrix in Eq.~(\ref{g2})
is set up for the complex energies along the contour, and the EMTO 
Green's function
is used to solve the electronic structure problem as outlined in Sec.~\ref{subsec_emto}. 
The partial waves $\phi_{Rl}(r_R)$ are obtained by radially integrating the Schr\"odinger
equation~(\ref{radialsch}) for the linearization energy $z = \epsilon_{Rl\nu}$. From these partial waves,
the LMTO potential parameters $C_{Rl}$ and $\Delta_{Rl}$ can be obtained (see Ref.~\onlinecite{skri.84}).
The LMTO Hamiltonian~(\ref{hamiltonian}) is constructed and diagonalized,
providing eigenvalues $\epsilon_j(\mathbf{k})$ and eigenvectors $u_{RL,j}(\mathbf{k})$. 
In the next step, the non-interacting local LMTO Green's function~(\ref{lmto_loc_green}) is computed for the
Matsubara frequencies $i\omega_n$. 
Correlation effects are generated by the interaction term, formally to be added to the non-interacting
Hamiltonian $H^\gamma$. The explicit form of the four index Coulomb interaction matrix elements is
discussed in Sec.~\ref{sec_res}.
From the Green's function formulated on the Matsubara axis the bath Green's function~(\ref{Gtilde}) at the LMTO level is obtained,
and passed into the DMFT many-body solver.

\begin{figure*}
\includegraphics[scale=0.5]{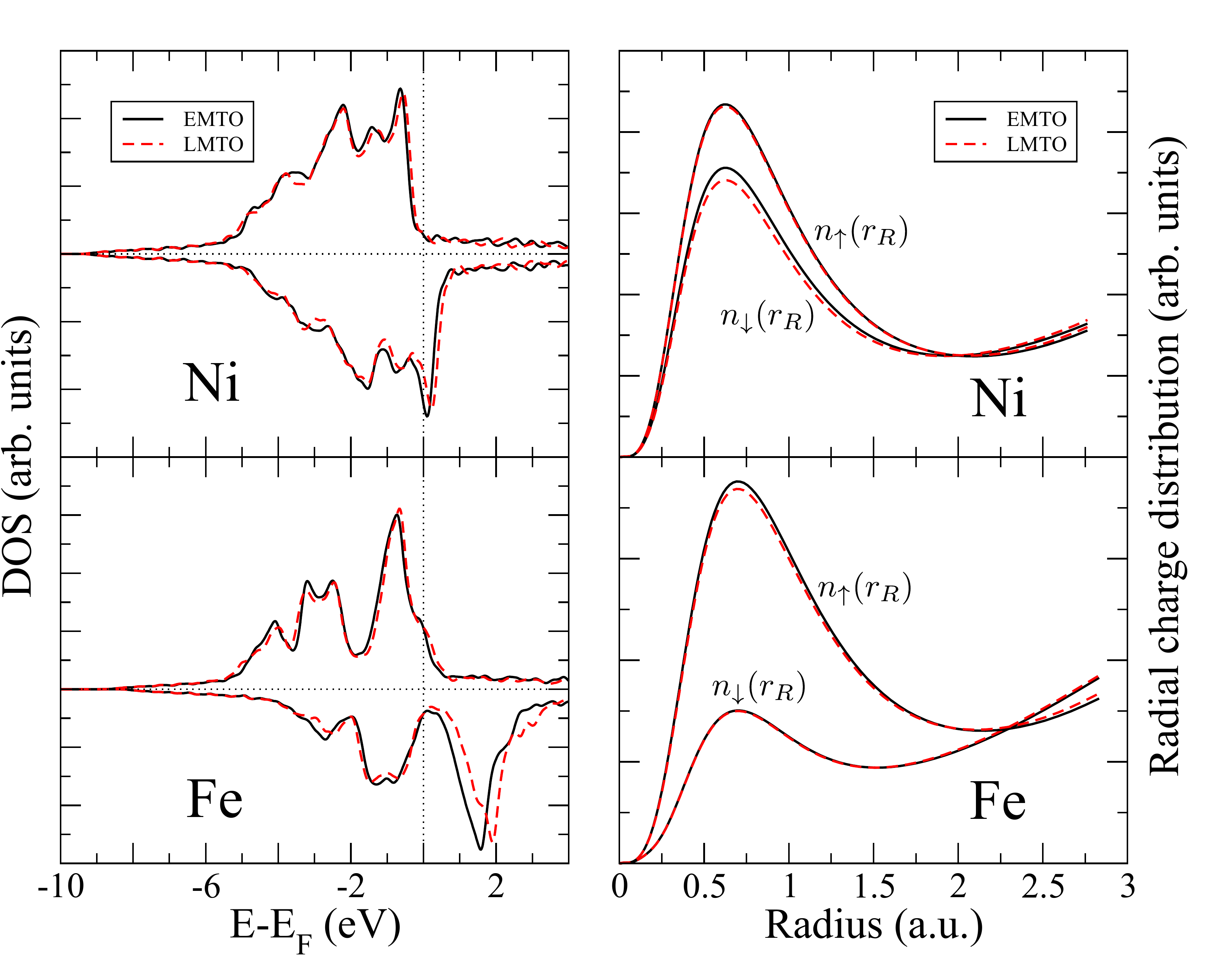}
\caption{(Color online) Left panel: Spin-resolved densities of states derived from EMTO (black solid line) and a 
linear approximation (red dashed line). 
(Top left) Majority and minority density of states of fcc Ni.
(Bottom left) Majority and minority density of states of bcc Fe.\\
Right panel: Spin-resolved valence electron charge density for Ni (top right) and Fe (bottom right). The EMTO charge is plotted
using black solid lines, while the charge stemming from linearization is shown with red dashed lines.}\label{lindos}
\end{figure*}

The error in the linearization can be assessed by comparing the density of states (DOS)
arising from the EMTO and the LMTO Green's functions, both at LDA level, see left panel of Fig.~\ref{lindos}.
The EMTO method was iterated self-consistently for Ni (above left) and Fe (below left),
using an $spd-$basis set. 
The DOS was then evaluated from the imaginary part of Eq.~(\ref{green}) (black solid lines)
and Eq.~(\ref{lmtogreen}) (red dashed lines), along a horizontal contour close to the real energy axis.
The curves are in good agreement with each other. The basis set linearization
will introduce approximations, but these are easily controlled and can in principle be improved by including higher order 
MTOs (the NMTO method~\cite{an.sa.00}).

\subsubsection{Elimination of $\Sigma(i\omega) \rightarrow \Sigma(z)$: \\ Charge density difference} \label{sig_pade}
An essential step for the charge self-consistency of LDA+DMFT with the EMTO-basis set~\cite{ch.vi.03}
is the analytic continuation of the self-energy $\Sigma_{RL,RL'}(i\omega)$ back to the complex contour,
which allows to update the path operator $g_{RL,RL'}(z)$ from which the real space charge density, Eq.~(\ref{emtocharge}),
is computed. The correlation effects upon the real space charge density has been analyzed in
the previous LDA+DMFT implementation for Fe, Ni and Cr~\cite{ch.vi.03}. In particular for Cr,
LDA+DMFT charge density shows accumulation of $d$ electrons due to correlation effects inside
the muffin-tin spheres and a depletion of density in the interstitial region. To capture
these correlation induced \emph{corrections} to the LDA charge density it seems natural for
the current implementation to propose the following scheme:
\begin{itemize}
\item The LDA charge density should be computed within EMTO, $n^{EMTO}_{LDA}({\bf r})$, on the complex contour,   
\item The DMFT charge density correction, $\Delta n^{\omega}({\bf r})$, should be computed within LMTO on the Matsubara axis.
\end{itemize}
To be specific, we outline the procedure: The LDA real space charge density 
is calculated from the complex contour, see Eq.~(\ref{emtocharge}).
Once the LMTO Green's function has been constructed on the Matsubara frequencies, the energy moments
Eqs.~(\ref{0mom})-(\ref{hmom}) are computed both on the LDA and the LDA+DMFT level.
This allows to evaluate the charge density $n^{LMTO}_{LDA(+DMFT)}({\bf r})$ 
according to Eq.~(\ref{lmtocharge}).
The charge density difference $\Delta n^{\omega}({\bf r})$ is then simply defined as
\begin{equation}\label{chargediff}
\Delta n^{\omega}({\bf r}) \equiv n^{LMTO}_{LDA+DMFT}({\bf r}) - n^{LMTO}_{LDA}({\bf r}),
\end{equation}
where the superscript of $\Delta n^{\omega}({\bf r})$ emphasize that this quantity is 
computed on the imaginary axis. The final LDA+DMFT real space charge density $n({\bf r})$ is 
obtained through
\begin{equation}\label{finalcharge}
n({\bf r}) \equiv n^{EMTO}_{LDA}({\bf r}) + \Delta n^{\omega}({\bf r}),
\end{equation}
and is used to close the self-consistent cycle. 
Note that the charges computed along the Matsubara axis contain contributions from all orbitals, and not only from the correlated subset.

To assess the possible differences between the EMTO and LMTO charge density, at the LDA level,
we plot in Fig.~\ref{lindos} (right column) the valence charge density for Ni/Fe in the upper/lower panel.
The EMTO charge densities (black solid lines) were iterated to self-consistency and evaluated according to Eq.~(\ref{emtocharge}).
The LMTO charge (red dashed lines) was evaluated from the EMTO self-consistent potentials by computing first
the energy moments of the LMTO Green's function Eq.~(\ref{lmtogreen}) using the contour integration~\cite{sk.ro.91},
and then applying Eq.~(\ref{lmtocharge}). The charge densities are in a very good agreement. 

\subsubsection{Total energy}
Within the Kohn-Sham scheme, the total energy functional can be expressed as
\begin{eqnarray}
E_{DFT}[n(\mathbf{r})] &=& T_s[n(\mathbf{r})] + 
\int \frac{n(\mathbf{r}')n(\mathbf{r})}{|\mathbf{r}'-\mathbf{r}|}\mathrm{d}\mathbf{r}'\mathrm{d}\mathbf{r}\nonumber\\
&&+ E_{xc}[n(\mathbf{r})] + \int V_{ext}n(\mathbf{r})\mathrm{d}\mathbf{r},
\end{eqnarray}
where $V_{ext}$ is the external ionic potential, $E_{xc}$ is the exchange-correlation
energy and $T_s$ is the Kohn-Sham single-particle kinetic energy. The square brackets indicate that the energy components are
\emph{functionals} of the density $n(\mathbf{r})$. For the proposed new method, the charge density
given as input is now computed on the LDA+DMFT level, Eq.~(\ref{finalcharge}), as outlined in the
previous section. A slight change in the expression of the kinetic energy:
\begin{eqnarray}
T_s[n(\mathbf{r})] &\equiv & \sum^{occ.}\limits_j \int \Psi_j(\mathbf{r})(-\nabla^2)\Psi_j(\mathbf{r})\mathrm{d}\mathbf{r}\nonumber \\
&=& \sum^{occ.}\limits_j \epsilon_j - \int n(\mathbf{r}) V_{eff}(\mathbf{r}) \mathrm{d}\mathbf{r},
\end{eqnarray}
has to reflect the change in the  one-electron energies $\epsilon_j$ caused
by the presence of the real part of the self-energy. Eq.~(\ref{kosh}) was
used for the second equality, of the above equation. In order to account for this 
change in the one-electron energies, the \emph{difference} between the LDA and LDA+DMFT
one-electron energies $\Delta \epsilon_j = \epsilon^{LDA+DMFT}_j - \epsilon^{LDA}_j$ is added to the expression for the kinetic energy. 
The total energy of a many-body system in the ground state includes also the Galitskii-Migdal contribution~\cite{fetter.walecka}.
This contribution is added into all LDA+DMFT computations. Other formulations such as the variational Luttinger-Ward functional
may give improved energies~\cite{sa.ko.04,ko.sa.06,ha.bi.15} but do not appear straightforward to implement in the present scheme.
In the current implementation the Galitski-Migdal energy contribution is computed
on the Matusbara axis in the LMTO formulation: 
\begin{eqnarray}
&E_{GM}& \equiv \nonumber\\
&\frac{T}{2} Tr_{{L}}&\sum\limits_n\sum\limits_{\mathbf{k}} G_{RL,R''L''}({\bf k},i\omega_n)\Sigma_{R''L'',R'L'}(i\omega_n)e^{i\omega_n0^+},
\end{eqnarray}
where $G_{RL,R''L''}({\bf k},i\omega_n)$ is on the LMTO LDA+DMFT level.
The final expression for the LDA+DMFT total energy is
\begin{eqnarray}\label{totenergy}
E_{LDA+DMFT}[n(\mathbf{r})] = E_{LDA}[n(\mathbf{r})] + \Delta \epsilon_j + E_{GM}
\end{eqnarray}
The Kohn-Sham $\epsilon_j$ one-electron energies from the DFT (LDA) calculation already
include some interaction effects through the Hartree and the exchange-correlation potential terms.
Including the interactions explicitly in the form of the Hubbard Hamiltonian, some interaction
contributions would be counted twice. Consequently, some double counting correction has to be included.
There is no universal solution to this problem, and most of the double counting schemes are empirical.
In the present method we take over the schemes used in the previous implementation~\cite{ch.vi.03}, a
detailed discussion is found in Ref.~\onlinecite{pe.ma.03}.

\subsection{Flow Diagram of the self-consistency calculation in $z$MTO+DMFT}

\begin{figure*}
\includegraphics[scale=0.8]{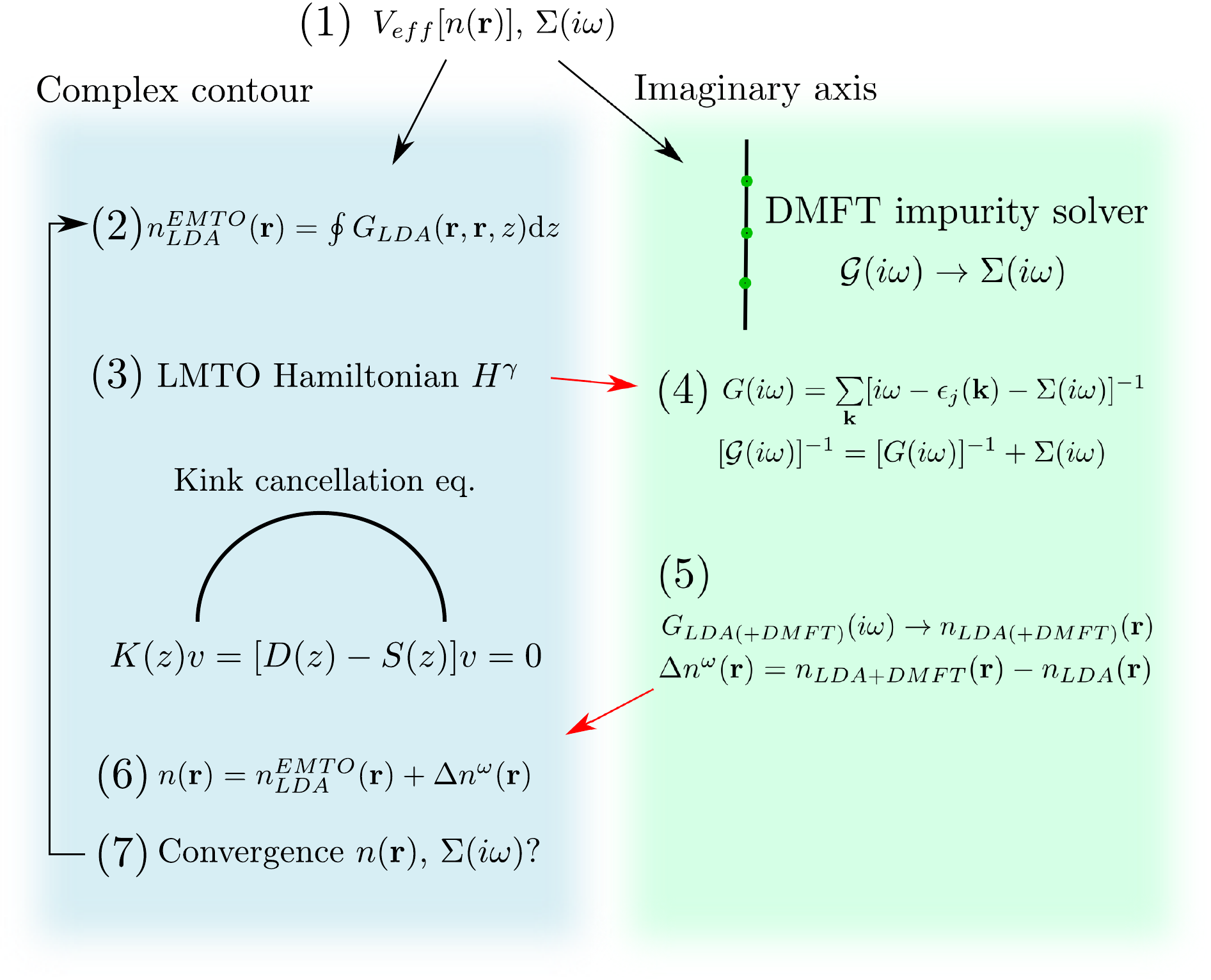}
\caption{(Color online) Schematic flow diagram of the new scheme. Note that within the cycle there is no
analytic continuation needed since the quantities passed between the complex contour and the imaginary
axis (red arrows) are energy independent (potential parameters and charge).}\label{flow}
\end{figure*}

The ideas presented in the previous section can be condensed in the following scheme 
that we call the $z$MTO+DMFT method (see Fig.~(\ref{flow})):
\begin{enumerate}[{(1)}]
\item The Kohn-Sham iterations are initiated
with a starting guess for the effective potential $V_{eff}(\mathbf{r})$ and the 
self-energy $\Sigma_{RL,RL'}(i\omega)$.
\item The kink-cancellation equations are constructed for points along the complex contour,
and the LDA level charge $n^{EMTO}_{LDA}({\bf r})$, Eq.~(\ref{emtocharge}), is obtained by 
integrating along the contour. At this stage, the LMTO potential parameters are also computed from the partial waves.
\item The Hamiltonian $H^{\gamma}$ is constructed from the potential parameters from step 
(2) using Eq.~(\ref{hamiltonian}), and the eigenvalue problem is solved.
\item The non-interacting LDA Green's function (LMTO) is constructed according to 
Eq.~(\ref{lmtogreenmats}) for the Matsubara frequencies from the Hamiltonian in step (3). 
The LMTO bath Green's function, Eq.~(\ref{Gtilde}), is computed and iterated into the 
DMFT self-consistency loop, from which a new $\Sigma_{RL,RL'}(i\omega)$ is obtained.
\item The $n^{LMTO}_{LDA(+DMFT)}({\bf r})$ charges are obtained by Matsubara summation, 
and the difference $\Delta n^{\omega}({\bf r})$ according to Eq.~(\ref{chargediff}) is evaluated.
\item The final LDA+DMFT charge density Eq.~(\ref{finalcharge}) is computed by adding 
$\Delta n^{\omega}({\bf r})$ from step (6) to the DFT charge density from step (2).
\item Return to step (2) until self-consistency in both the charge and self-energy is reached.
\end{enumerate}
Once the self-consistency has been reached, observables such as the total energy Eq.~(\ref{totenergy}) and
spectral functions can be evaluated. Note that the spectral functions are evaluated on a horizontal
contour slightly shifted away from the real axis. To analyze the self-energy along the real axis, a Pad\'{e} approximant
is can be used. Note however, that this does not affect the Kohn-Sham
loops, and has to be carried out only once at the end, after self-consistency has been reached. 
In this case is also easy to identify spurious poles in the Pad\'{e} approximant, as outlined in Ref.~\onlinecite{os.ch.12}.

\section{Results}\label{sec_res}

\begin{figure*}
\includegraphics[scale=0.35,clip=true]{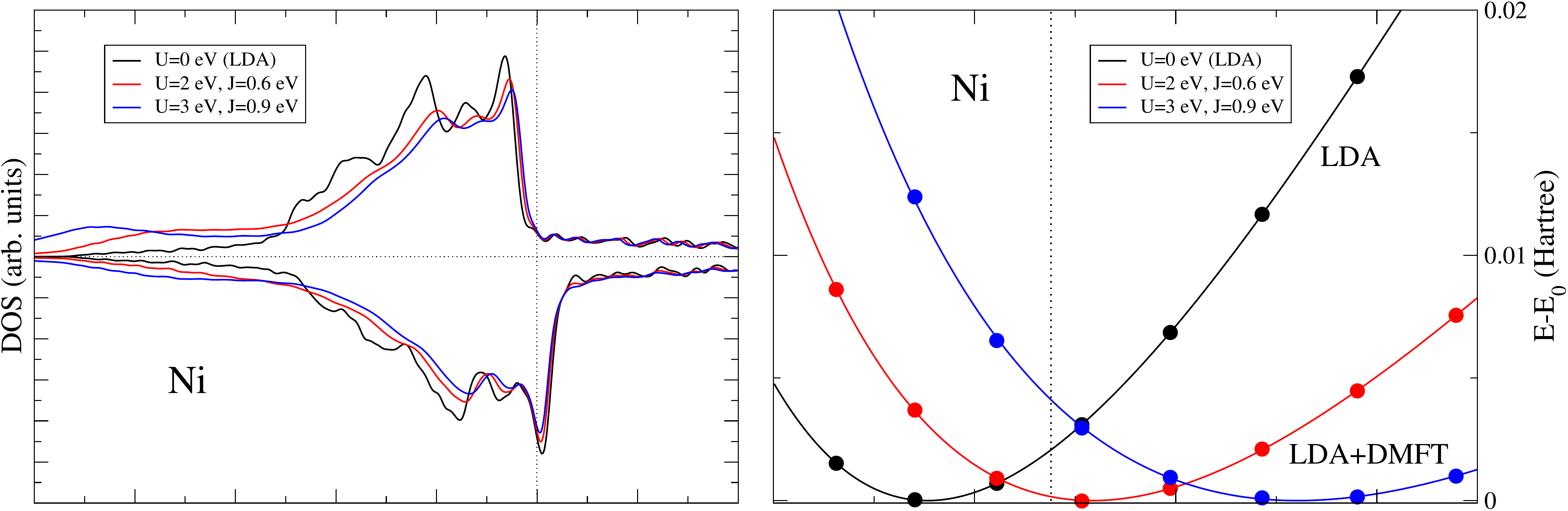}\\
\includegraphics[scale=0.35,clip=true]{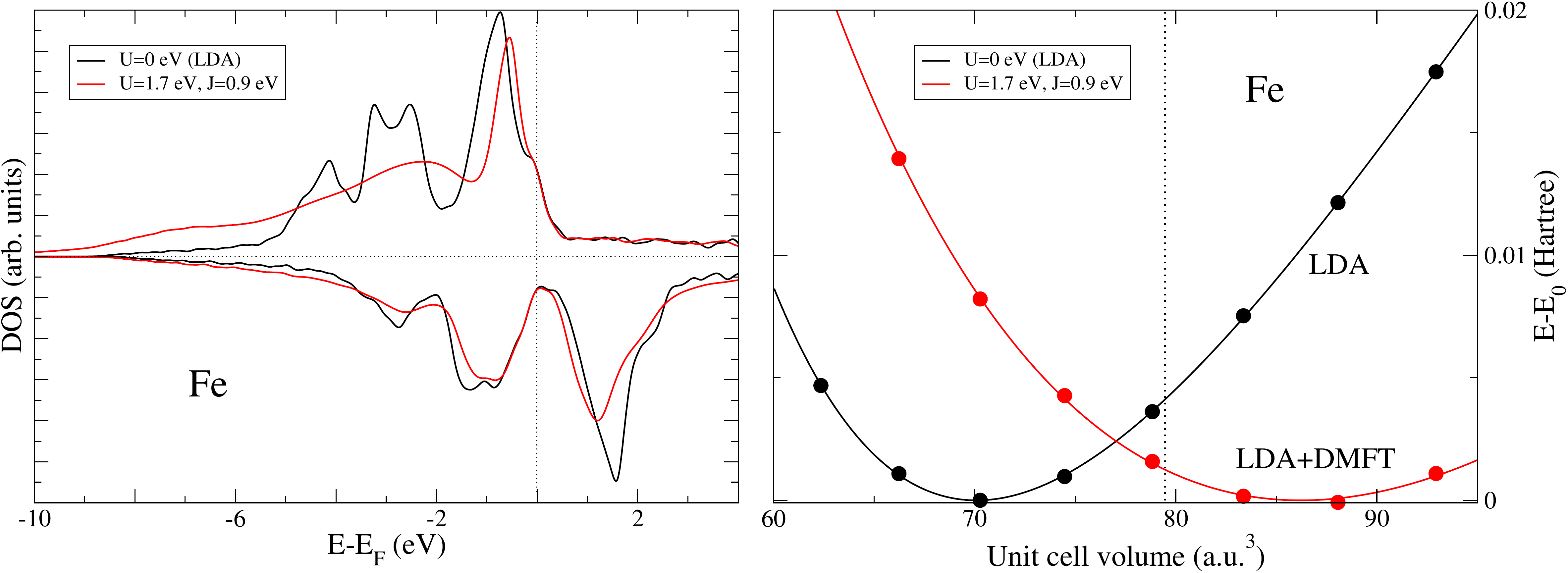}
\caption{(Color online) Spin-resolved density of states (left) and equation of state (right) for nickel (top) and
iron (bottom), for various values of the
Coloumb parameters $U$ and $J$. For Ni, the DOS was calculated for a unit cell volume 73.79 a.u.$^3$, and for Fe
the unit cell volume was 78.84 a.u.$^3$. In the right panel the dotted lines marks the experimental volumes.}\label{ni}
\end{figure*}

To assess the implementation electronic structure calculations have been performed according to the method proposed above. 
Transition metals and compounds in which  the $d$-orbitals form the correlated basis set have been considered.
For the DMFT impurity solver a fluctuation exchange (FLEX)~\cite{bi.sc.89} 
type of approximation was used for the multiorbital case~\cite{li.ka.98,ka.li.99,po.ka.05}.
In contrast to the original formulation of                                          
FLEX~\cite{bi.sc.89}, the spin-polarized $T$-matrix FLEX (SPTFLEX), used for
 the present calculations treats the particle-particle and the particle-hole channel
differently~\cite{li.ka.98,ka.li.99,po.ka.05}. While the particle-particle
processes are important for the renormalization of the effective
interaction, the particle-hole channel describes the interaction
of electrons with the spin-fluctuations. In addition the advantage                                    
of such a computational scheme is that the electron-electron interaction term can be
considered in a full spin and orbital rotationally invariant form, viz.
$\frac{1}{2}\sum_{{i \{m, \sigma \} }} U_{mm'm''m'''}
c^{\dag}_{im\sigma}c^{\dag}_{im'\sigma'}c_{im'''\sigma'}c_{im''\sigma} $.
Here, $c_{im\sigma}/c^\dagger_{im\sigma}$ annihilates/creates an electron with 
spin $\sigma$ on the orbital $m$ at the lattice site $i$.
The Coulomb matrix elements $U_{mm'm''m'''}$ are expressed in the usual
way~\cite{im.fu.98} in terms of Slater integrals.
Since specific correlation effects are already
included in the exchange-correlation functional, so-called
``double counted'' terms must be subtracted. To achieve this, we replace
$\Sigma_{\sigma}(E)$ with $\Sigma_{\sigma}(E)-\Sigma_{\sigma}(0)$
\cite{li.ka.01} in all equations of the DMFT procedure \cite{ko.sa.06}.
Physically, this is related to the fact that DMFT only adds {\it dynamical}
correlations to the DFT result~\cite{pe.ma.03}.

\subsection{Transition metals: nickel and iron}

Within the family of the late $3d$ transition metals, nickel and iron are known to show in their band structures signatures of electronic correlation~\cite{li.ka.01}. 
Nickel is well-known for a ``6-eV-satellite'' in its photoemission spectra~\cite{liebsch.79}, while a similar satellite in
iron is debated~\cite{gr.ma.07,sa.fi.09}. 

For both Fe and Ni, a $spd$-basis was used, and the $4s$ and $3d$ states were treated as valence. The
core electron levels were recalculated at each Kohn-Sham iteration (soft-core approximation). The
kink cancellation condition was set up for 16 energy points distributed around a semi-circular
contour with a diameter of 1 Ry, enclosing the valence band. The BZ integrations
were carried out on an equidistant mesh with 285 \textbf{k}-points (for Fe) and 240 \textbf{k}-points (for Ni) 
in the irreducible BZ. For the exchange-correlation potential the 
local spin density approximation with the Perdew-Wang parameterization~\cite{pe.wa.92} was used.
For the DMFT impurity calculations, the Matsubara frequencies were truncated after 2048 frequencies, and the temperature
was set to $T=400$ K. The values for the Coulomb $U$ and the exchange $J$ parameters are
discussed in connection with the presentation of the results in each case. 
The equations of state were obtained by fitting the energy-versus-volume data to a Birch-Murnaghan function~\cite{birch.47}.
The densities of state were computed along a horizontal contour shifted a distance $\delta=0.02$ Ry
away from the real energy axis. At the end of the selfconsistent calculations, to obtain the self-energy on a
real energy mesh, $\Sigma(\omega)$ can be analytically continued into a horizontal contour by a Pad\'e approximant
constructed by the Thiele method~\cite{vi.se.77}.

\begin{table*}
\caption{Computed equilibrium volumes $V_0$ (a.u.$^3$) and bulk modulus $B_0$ (GPa) for fcc Ni and bcc Fe. Comparison
is made with theoretical and experimental references. Data in parenthesis next to a quantity is the relative difference 
between the quantity and the LDA ($U=0$) value, $\delta x \equiv (x-x_{LDA})/x_{LDA}$. Experimental
data taken from Ref.~\onlinecite{young}.}\label{eostab}
\begin{ruledtabular}
\begin{tabular}{lll@{\hspace{-20mm}}ll@{\hspace{-20mm}}ll}
\textbf{Ni} &  LDA & $U=2$ eV & & $U=3$ eV & & Exp.\\
& $V_0$ & $V_0$ & $\delta V_0$ & $V_0$ & $\delta V_0$ &\\
\hline
This work & 67.65 & 75.84 & (0.12)  & 86.04 & (0.27) &\\
 FP-LMTO (Ref.~\onlinecite{ma.mi.09}) & 67.88 & 76.20 & (0.12) & 89.48 & (0.31) & 73.79 \\
 KKR (Ref.~\onlinecite{ma.mi.09}) & 66.86 & 76.28 & (0.14) & 85.53 & (0.28) &\\
 & & & & & &\\
 & $B_0$ & $B_0$ & $\delta B_0$ & $B_0$ & $\delta B_0$ &\\
\hline
 This work & 259 & 162 & (-0.37) & 99 & (-0.62) &\\
 FP-LMTO (Ref.~\onlinecite{ma.mi.09}) & 260 & 163 & (-0.37) & 84 & (-0.68) & 179\\
 KKR (Ref.~\onlinecite{ma.mi.09}) & 280 & 171 & (-0.39) & 132 & (-0.53) &\\
 & & & & & &\\
\hline
\hline
\textbf{Fe} & LDA & $U=1.7$ eV & & & &Exp.\\
& $V_0$ & $V_0$ & $\delta V_0$ & & &\\
\hline
 This work & 70.09 & 86.21 & (0.23) & & \\
 FP-LMTO (Ref.~\onlinecite{gr.ma.12}) & 70.49 & 87.06 & (0.24) & & & 79.46 \\
  & & & & & &\\
 & $B_0$ & $B_0$ & $\delta B_0$ & & &\\
\hline
 This work & 253 & 124 & (-0.51) & &\\
 FP-LMTO (Ref.~\onlinecite{gr.ma.12}) & 234 & 90 & (-0.62) & & & 163 \\
\end{tabular}
\end{ruledtabular}
\end{table*}

In the top left part of Fig.~\ref{ni}, the LDA and LDA+DMFT density of states for Ni is presented. 
The volume was set to the experimental value (73.79 a.u.$^3$). The
new method compares well with previous DFT+DMFT studies employing the 
SPTFLEX impurity solver~\cite{ch.vi.03,mi.ch.05,gr.ma.07}, and captures the main correlation effects of Ni
such as the satellite formation and band narrowing. Note that the correlation effects are stronger
in the majority spin channel (more pronounced satellite, more narrow bandwidth) than in the minority
spin channel, which is common for the late $3d$ metals. For the case of $U=3$ eV (blue line), the position of 
the ``6-eV'' satellite is at higher binding energy than in experiment. The value $U=3$ eV has previously
given the correct satellite position when a quantum Monte Carlo impurity solver was used~\cite{li.ka.01},
and the fact that the SPTFLEX solver overestimate the effect of correlation is thought to be due to the
perturbative nature of the solver~\cite{ka.li.02}.  Recent spin-polarized positron annihilation experiments and
LDA+DMFT calculations allowed to determine the value for the local electron-electron interaction strength in
ferromagnetic nickel to the value of $2 \pm 0.1$ eV~\cite{ce.we.16}.
By decreasing the Coloumb parameter to $U=2$ eV (red line),
the satellite is shifted to lower binding energy, in better agreement with experiment, as found previously~\cite{ka.li.02}.

The top right part of Fig.~\ref{ni} shows the equation of state of Ni as calculated within the new
method, for various values of the Coulomb parameters $U$ and $J$. 
The effect of correlation can be seen to increase the equilibrium volume
from the value given by the LDA (corresponding to $U=0$, black line). The equilibrium volumes
are given in Table~\ref{eostab}, together with the bulk moduli. As already mentioned in the discussion
of the nickel DOS, the SPTFLEX solver overestimates the effect of correlation. This is seen for
the equilibrium volume, where the commonly accepted value of $U=3$ eV (blue line) overestimates
the equilibrium volume. $U=2$ eV (red line) gives a better agreement with the experimental volume.
It should also be noted that the bulk modulus is softened as $U$ is increased, which corrects for the
overestimation made by the LDA functional.

Fig.~\ref{ni} shows the DOS (bottom left) and equation of state (bottom right) for bcc Fe, 
for the case of standard LDA ($U=0$) and for $U=1.7$ eV, $J=0.9$ eV. 
Similar values of $U$ and $J$ have previously been successfully used to describe the photoemission
spectra and energetics of iron~\cite{sa.fi.09,le.po.11,gr.ma.12}.
The effect of correlation is seen to broaden the peaks in the DOS, and create a satellite structure
at $\sim 7$ eV binding energy, in agreement with previous SPTFLEX studies~\cite{gr.ma.07,gr.ma.12}. 
By including local correlation effects, the equilibrium volume is increased, similar as for Ni.
This can be seen in the bottom right part of Fig.~\ref{ni}, where the equation of state is given.
The effect of correlation also reduces the bulk modulus (see Table~\ref{eostab}). The agreement
between our results and the ones from the Ref.~\onlinecite{gr.ma.12} is very good, the slight differences are
due to the spin-orbit coupling explicitly present in Ref.~\onlinecite{gr.ma.12}. On the other hand it is known
that spin-orbit effects are quite small for Fe \cite{er.jo.90}.

\subsection{Iron aluminium}

\begin{figure*}
\includegraphics[scale=0.35]{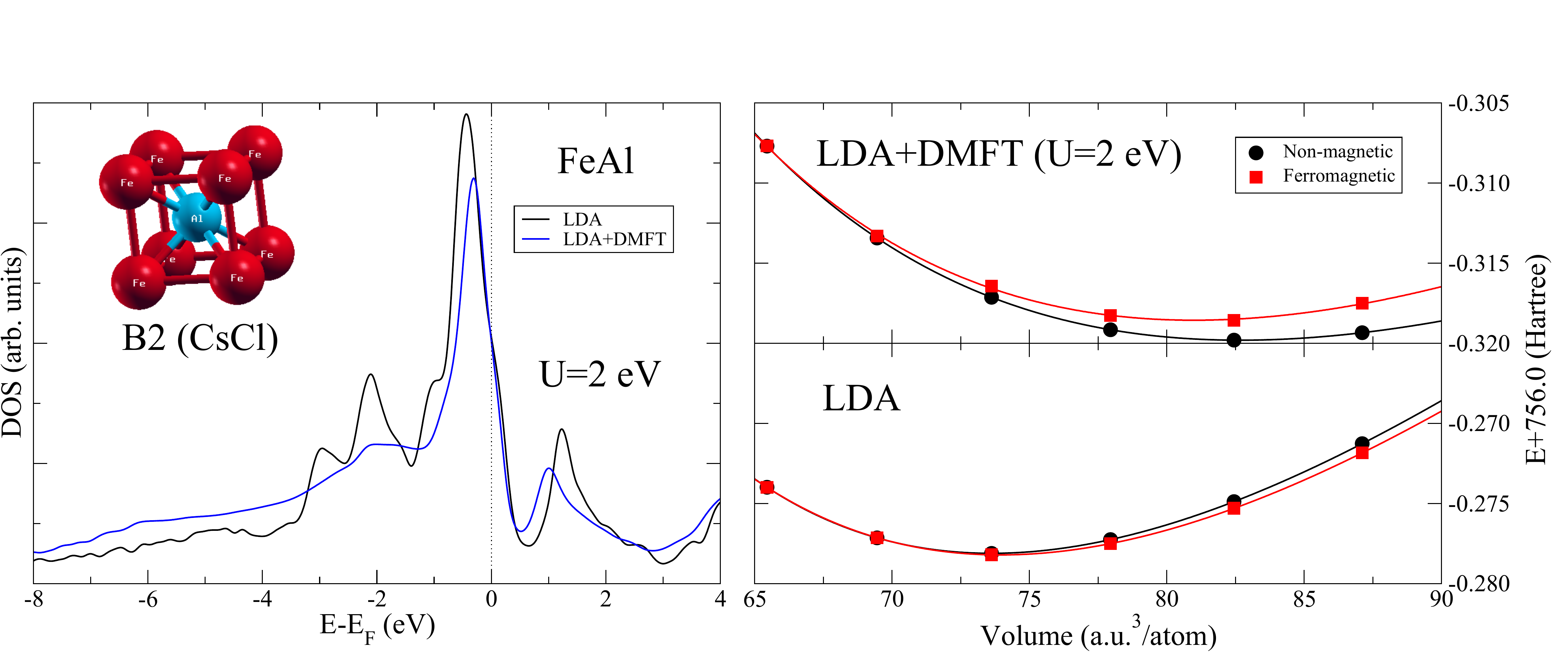}
\caption{Density of states (left) and equations of state (right) of FeAl.}\label{feal}
\end{figure*}

The stoichiometric intermetallic compound FeAl has attracted the interest of the electronic structure
community mainly due to its magnetic properties. 
While FeAl is paramagnetic in experiment, LSDA calculations within density functional theory predict an
ordered ferromagnetic ground state  with a magnetic moment of about $\sim 0.7$ $\mu_B$. 
Mohn et al.~\cite{mo.pe.01} showed that including the effect of the local Coulomb interaction $U$ through
the LDA+U method the nonmagnetic
state can be stabilized for a narrow range of $U$ values. It was further argued that the reduction in the
DOS at the Fermi level, caused by increasing $U$ values, will favor the nonmagnetic state through the Stoner criteria.
Petukhov et al.~\cite{pe.ma.03} pointed out the importance of dynamic effects
by LDA+DMFT calculations of the spectral functions, showing that the nonmagnetic solution is stable within
LDA+DMFT, and that the DOS is pinned to the Fermi level. Later on, Galler et al.~\cite{ga.ta.15} confirmed this, 
while also computing susceptibilities for FeAl within LDA+DMFT using a continuous-time quantum Monte Carlo (CT-QMC)
impurity solver. None of the above previous LDA+DMFT studies presented total energies.

We have investigated the electronic structure of FeAl with our new method, in order to evaluate the density of states
and the total energy for volumes around the experimental value. 
FeAl crystallizes in the B2 (CsCl) structure, i.e. a simple cubic lattice with Fe at position ($0,0,0$)
and Al at ($\frac{a}{2},\frac{a}{2},\frac{a}{2}$), where the experimental lattice constant is $a=5.496$ a.u.~\cite{mo.pe.01}
(note that also the value $a=5.409$ a.u. is reported in the literature~\cite{su.sa.95,ku.po.99}). 
An $spd$-basis was used, and a contour of diameter 1 Ry with 16 energy points was employed for the energy integrations.
For the BZ integration 286 \textbf{k}-points in the irreducible part was employed.

In the left part of Fig.~\ref{feal} we present the non-magnetic density of states for FeAl, computed assuming
$U=0$ eV (black line) and $U=2$ eV (blue line).
As pointed out in the previous studies~\cite{pe.ma.03,ga.ta.15}, the increasing of the Coulomb $U$ parameter,
within LDA+DMFT has little effect on the density of states at the Fermi level, in contrast to LDA+U calculations~\cite{mo.pe.01},
while it leads to a band narrowing. This is an indication that spin-fluctuations, which are included on a perturbative
level in the SPTFLEX solver, changes the simple picture of Stoner instability.

In the right panel of Fig.~\ref{feal}, our computed total energies for ferro- and non-magnetic FeAl are presented.
In the case of LDA ($U=0$, bottom right), the non-magnetic total energy (black line) is never lower than the
ferromagnetic total energy (red line), for all the studied volumes. 
In the lower volume range the ferromagnetic moment is lost, indicated by the coincidence of the two energy curves
$\lesssim 70$ a.u.$^3$. The fact that a ferromagnetic ground state is favored in LDA is in agreement with previous DFT
studies~\cite{ku.po.99}. The equilibrium volumes for the respective curves
are 73.95 a.u.$^3$ ($a=5.288$ a.u.) for the ferromagnetic curve, and 73.62 a.u.$^3$ ($a=5.280$ a.u.) for 
the non-magnetic curve, and hence the ferro- and non-magnetic lattice constants differ by $<1\%$ only.
Previous DFT studies have found lattice constants of value $a=5.397$ a.u. (TB-LMTO, non-local corrections to the LDA,
Ref.~\onlinecite{ku.po.99}), 
$a=5.364$ a.u. (TB-LMTO, Barth-Hedin parametrization of LDA, Ref.~\onlinecite{su.sa.95}) and $a=5.330$ a.u.
(full-potential linearized augmented Slater-type orbital method using LDA, Ref.~\onlinecite{wa.we.98}),
using different basis sets and exchange-correlation functionals. The previously reported lattice constants are all
larger than the current results, but are consistent given the fact that different basis sets and exchange-correlation
functionals were used.

As local correlation effects are taken into account within LDA+DMFT ($U=2$ eV, top right), the situation
is reversed. In this case the ferromagnetic solution is always higher in energy compared to the
non-magnetic solution, indicating that the non-magnetic solution is the ground state for the whole
volume range. For volumes $\lesssim 67$ a.u.$^3$, the magnetic moment is lost, and the two curves
coincide. The equilibrium volumes for the respective curves
are 80.99 a.u.$^3$ ($a=5.451$ a.u.) for the ferromagnetic curve, and 82.67 a.u.$^3$ ($a=5.489$ a.u.) for 
the non-magnetic curve, which is in good agreement with experiment.

Associating the analysis of the DOS and equation of state, we see that LDA+DMFT is able to
explain the experimentally observed fact that FeAl is in a non-magnetic ground state, while
at the same time providing an equilibrium lattice constant in better agreement with experiment
than the LDA. By investigating the DOS, the Stoner criterion (an increased DOS at $E_F$ is leading to
a magnetic instability) for ferromagnetism can be ruled out as an explanation for the
magnetism in FeAl.

\section{Conclusion and Outlook}\label{sec_conc}
In this paper we have introduced a new computational scheme for LDA+DMFT calculations, using Green's function methods.
The new method is able to describe correlated systems such as transition metals and compounds, and shows
results in very good agreement with previous LDA+DMFT implementations.
At the heart of the current implementation is the formulation of the LDA Green's function directly on the Matsubara axis,
using the Lehmann representation in terms of the eigenvalues and eigenfunctions of the LMTO Hamiltonian.
This simple procedure is essential for circumventing the analytical continuation of the Green's function from the
complex contour to the Matsubara frequencies (Sec.~\ref{g_pade}). The real advantage of this construction appears
in the computation of the charge density. Starting from the zeroth moment of the LMTO Green's function, the extension
to higher order moments becomes possible. 
From these moments the real space charge can be constructed. The  difference between correlated
and non-correlated charge density allows for the self-consistency and in the same time circumvent
the second analytical continuation, that of the self-energy from the Matsubara axis to the complex contour (Sec.~\ref{sig_pade}).
The idea to consider charge density differences between LDA and LDA+DMFT might also prove useful for
Hamiltonian based methods, since the operation of subtraction could help in reducing systematic errors
coming from the numerically difficult Matsubara sums.

By side-stepping the ill-posed analytic continuation problems, a numerically stable implementation is
possible, at the minor cost of performing basis set linearization for the calculations along the imaginary axis.    

Numerical results are presented for Fe and Ni. A direct numerical comparison between the imaginary part of
the EMTO and the LMTO Green's functions along a horizontal contour close to the real axis is studied in Fig.~\ref{lindos}.
The agreement between the basis sets as well as for radially distributed real space charge are found to be excellent.
The $z$MTO+DMFT densities of states and total energy curves are then presented in Fig.~\ref{ni}, and are found to be
in very good agreement with previous LDA+DMFT studies that were employing other basis sets. 
As a final example, the spectral functions and equations of state of the FeAl transition metal compound is studied.
Similarly, an excellent agreement is found when comparing to previous LDA+DMFT methods~\cite{pe.ma.03}. For a Coulomb
interaction strength of magnitude $U=2$ eV (on Fe in FeAl), the total energies for FeAl are seen to favor a non-magnetic
ground-state, in accordance with experiment.

As an outlook, we propose several possibilities to extend the current $z$MTO+DMFT implementation. 
First, the  downfolding of the linearized basis set can be included~\cite{po.am.07},
in order to reduce the size of the minimal basis set even further.
Second, the full-charge density (FCD) technique~\cite{vi.ko.94} applied to the EMTO method has previously provided accurate
total energies for low-symmetry structures, while still keeping the efficiency of the
spherical potential approximation (see Ref.~\cite{vitos.10}
The implementation of the FCD into the $z$MTO+DMFT method would make it possible to study the energetics
of low-symmetry structures and anisotropic lattice distortions of correlated materials, which currently is work in progress.
Finally, a major motivation is to enable a combination of the present method with the coherent-potential approximation~\cite{soven.67},
or with the typical medium theory for disorder~\cite{te.zh.17}. This would provide a method that could handle strong correlation
and disorder in alloy systems, including the problem of Anderson localization~\cite{te.zh.17}.

In conclusion we have attempted to demonstrate by means of elementary examples that the current
$z$MTO+DMFT, in conjunction with the perturbative SPTFLEX solver, can successfully describe the electronic
structure and energetics of transition metals and their compounds.
Even though the SPTFLEX solver is numerically simple due to its algebraic structure, it is still sufficiently rigorous
to deal with correlated electrons in condensed matter. A more sophisticated implementation using a variant
of Continuous Time  Quantum Monte-Carlo, DMFT impurity solvers is in progress. 

\section*{Acknowledgments}
We greatly benefited from the discussions with D. Vollhardt and O. K. Andersen,
whose advices are gratefully acknowledged.  
Financial support of the Deutsche Forschungsgemeinschaft through FOR 1346 is gratefully acknowledged.
A.\"{O}. acknowledges helpful discussions with I. Di Marco.
L.V.  acknowledges financial support from the Swedish Research Council,
the Swedish Foundation for Strategic Research, the Swedish
Foundation for International Cooperation in Research and
Higher Education, and the Hungarian Scientific Research Fund
(OTKA 84078 and 109570).
We acknowledge computational resources provided by the Swedish National Infrastructure for Computing (SNIC) 
at the National Supercomputer Centre (NSC) in Link\"{o}ping.

\bibliography{references_database,references_lemto}

\end{document}